\documentclass[trackchanges]{aastex701}
\usepackage{xcolor}
\begin{document}

\title{Discovery of activity on Main Belt Asteroid 6993 (1995 BJ4)}

\author[orcid=0009-0008-0758-9192,gname=Erin, sname='Clark']{Erin M. Clark}
\affiliation{Department of Planetary Science and Astronomy, Northern Arizona University, Flagstaff, AZ 86011, USA}
\email[show]{emc635@nau.edu}  

\author[orcid=0000-0003-4580-3790,gname=David, sname='Trilling']{David E. Trilling} 
\affiliation{Department of Planetary Science and Astronomy, Northern Arizona University, Flagstaff, AZ 86011, USA}
\email{@david.trilling@nau.edu}

\author[orcid=0009-0006-2731-5522,gname=Revanth, sname='Munugala']{Revanth Munugala} 
\affiliation{School of Informatics, Computing, and Cyber Systems, Northern Arizona University, Flagstaff, AZ 86011, USA}
\email{@email.com}

\author[orcid=0000-0002-6702-7676,gname=Michael, sname='Kelley']{Michael S.P. Kelley} 
\affiliation{Department of Astronomy, University of Maryland, College Park, MD 20740, USA}
\email{@email.com}

\author[orcid=0000-0001-7225-9271,gname=Henry, sname='Hsieh']{Henry Hsieh} 
\affiliation{Planetary Science Insitute, Tuscon, AZ 85719, USA}
\email{@email.com}

\author[orcid=0000-0002-0826-6204,gname=Michael, sname='Gowanlock']{Michael Gowanlock} 
\affiliation{School of Informatics, Computing, and Cyber Systems, Northern Arizona University, Flagstaff, AZ 86011, USA}
\email{@email.com}

\author[orcid=0009-0005-9955-1500,gname=Andrew, sname='McNeill']{Andrew McNeill} 
\affiliation{Department of Physics and Astronomy, Bowling Green State University, 806 Ridge Street, Bowling Green, OH 43403, USA}
\email{@email.com}

\author[orcid=0009-0006-9247-575X,gname=Savannah, sname='Chappus']{Savannah Chappus} 
\affiliation{School of Informatics, Computing, and Cyber Systems, Northern Arizona University, Flagstaff, AZ 86011, USA}
\email{@email.com}

\author[orcid=0000-0002-6292-9056,gname=Maria, sname='Chernyavskaya']{Maria Chernyavskaya} 
\affiliation{Department of Astronomy, University of Maryland, College Park, MD 20740, USA}
\email{@email.com}

\author[gname=Graceson, sname='Mule']{Graceson Mule} 
\affiliation{School of Informatics, Computing, and Cyber Systems, Northern Arizona University, Flagstaff, AZ 86011, USA}
\email{@email.com}

\author[orcid=0009-0008-0547-3962,gname=Gunnar, sname='Johnson']{Gunnar Johnson} 
\affiliation{Department of Planetary Science and Astronomy, Northern Arizona University, Flagstaff, AZ 86011, USA}
\email{@email.com}

\author[orcid=0000-0002-6676-1713,gname=Daniel, sname='Kramer']{Daniel Kramer} 
\affiliation{School of Informatics, Computing, and Cyber Systems, Northern Arizona University, Flagstaff, AZ 86011, USA}
\email{@email.com}

\begin{abstract}

We report the discovery of activity on asteroid 6993 (1995 BJ4) in the Zwicky Transient Facility (ZTF) alert stream data on UT 2026~August~5, with the help of machine learning. Visible activity can be seen as early as UT 2026~June~25. Activity on this outer-main belt object is likely not primarily
driven by rotational instability, but rather is consistent with
an impact event or sublimation based on several arguments. If this activity is sublimation driven, 6993 would be the largest known main belt comet (11.85 km), with the largest perihelion distance (2.853 au) and an unusually high geometric albedo of 0.166 ± 0.030. Asteroid 6993 should be monitored as it approaches perihelion in 2027~January to distinguish between activity mechanisms and understand its potential volatile inventory.

\end{abstract}



\section{Introduction} 
Active asteroids are a rare subset of the asteroid population that exhibit comet-like behaviors, displaying comae or tails. Activity (i.e., comet-like mass loss) on these exotic objects is generally transient and driven via rotational spin-up, impacts, or volatile sublimation. Those that are sublimating can be classified as Main Belt Comets (MBCs; \citealt{Hsieh_Jewitt_2006}) and often become active with proximity to the Sun. Recent James Webb Space Telescope NIRSpec observations reveal that outer main belt MBCs appear to be driven by water-ice sublimation and are depleted in CO$_2$ relative to other comets \citep{Kelley_2023, Hsieh_2025, Hsieh_2026}, implying that MBCs are representative of a sample of volatile material that is underrepresented in observations of classical comets. The classical comets formed in the outer Solar System and therefore present different compositions than the MBC population, which likely formed in the asteroid belt. Because of this, understanding MBCs may
directly inform our estimates of the volatile inventory of the Solar
System, particularly in the inner Solar System, as well as the potential roles these objects could have had in delivering materials crucial to the origins of life on Earth \citep{Meech_Raymond_2019}.

There are currently fewer than 100 known active asteroids, with an estimated occurrence rate among main-belt asteroids being $\sim$1 in 10,000 \citep{Hsieh_2015, Jewitt_2015, Chandler_2018}. This small sample size limits population-level studies of activity mechanisms and prevents robust constraints on the abundance of water in the asteroid belt. To overcome these limitations, we turn to large-scale all-sky surveys, such as the Zwicky Transient Facility (ZTF; \citealt{Bellm_2019, Masci_2019}) and the Vera C. Rubin Observatory’s Legacy Survey of Space and Time (LSST; \citealt{Ivezic_2019}). LSST is expected observe $\sim$5~million asteroids \citep{Kurlander_2025}, compared to the $\sim$1~million known, presenting data volume challenges never before faced by the Solar System community. In preparation for the data rates produced by LSST, we have built and applied automated active asteroid candidate detection tools within the framework of the Solar System Notification Alert Processing System (SNAPS; \citealt{SNAPS, Trilling_2026}). We have been using the ZTF alert stream as a development testbed for these tools and as an independent experiment, and it is the application of these tools that allowed us to discover the activity on main belt asteroid 6993 that we present here.

Asteroid 6993 has 528 ZTF observations ingested by SNAPS from UT 2019~May~4 to 2026~September~22. Asteroid 6993 is an outer-main belt asteroid with a
semi-major axis of a $=$ 3.108 au, an aphelion distance of Q $=$ 3.363 au, and
a perihelion distance of q~$=$~2.853 au. With an absolute magnitude of $H_V=$ 12.43 (diameter around D $=$ 11.85 km; \citealt{NEOWISE}), asteroid 6993 has a relatively low inclination and eccentricity of i $=$ 2.4 degrees and e $=$ 0.08. Active asteroids are distinguishable from Jupiter family comets by being dynamically asteroidal with a Tisserand parameter of $T_J>$ 3.08 \citep{Jewitt_Hsieh_2024}, and 6993 fits this description with $T_J=$ 3.2.

In Section 2, we briefly discuss the candidate identification and
confirmation methods that enabled the discovery of activity on
asteroid 6993. In Section 3, we present our activity detections and in Section 4, we address the possible activity mechanisms driving the observed emission on asteroid 6993. Finally, we present some future directions in Section 5.

\section{Detection of Activity}
Automated candidate detection tools are critical when dealing with
large data volumes, now reaching up to 20~TB/night for all celestial sources \citep{Ivezic_2019}. We apply two methods to identify active asteroid
candidates: \textbf{(1)} Image classification using a machine learning model in the form of a convolutional neural network (CNN) to identify sources appearing to have comae or tails, and \textbf{(2)} point spread function (PSF) comparison to nearby sources imaged close in time to identify extended sources with comae. Details of these algorithms will be presented in a subsequent paper (Clark et al., in prep.).
\begin{figure}[ht!]
\centering
    \includegraphics[width=0.75\textwidth]{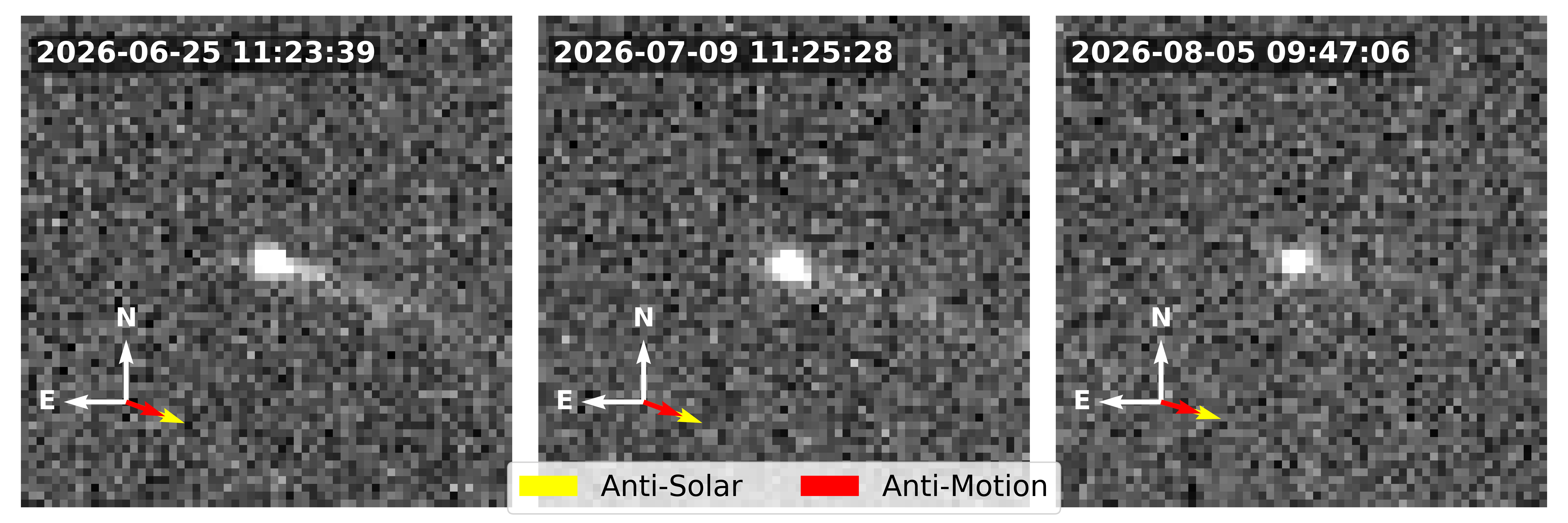}
    
    \vspace{0.3cm} 
    
    \includegraphics[width=0.50\textwidth]{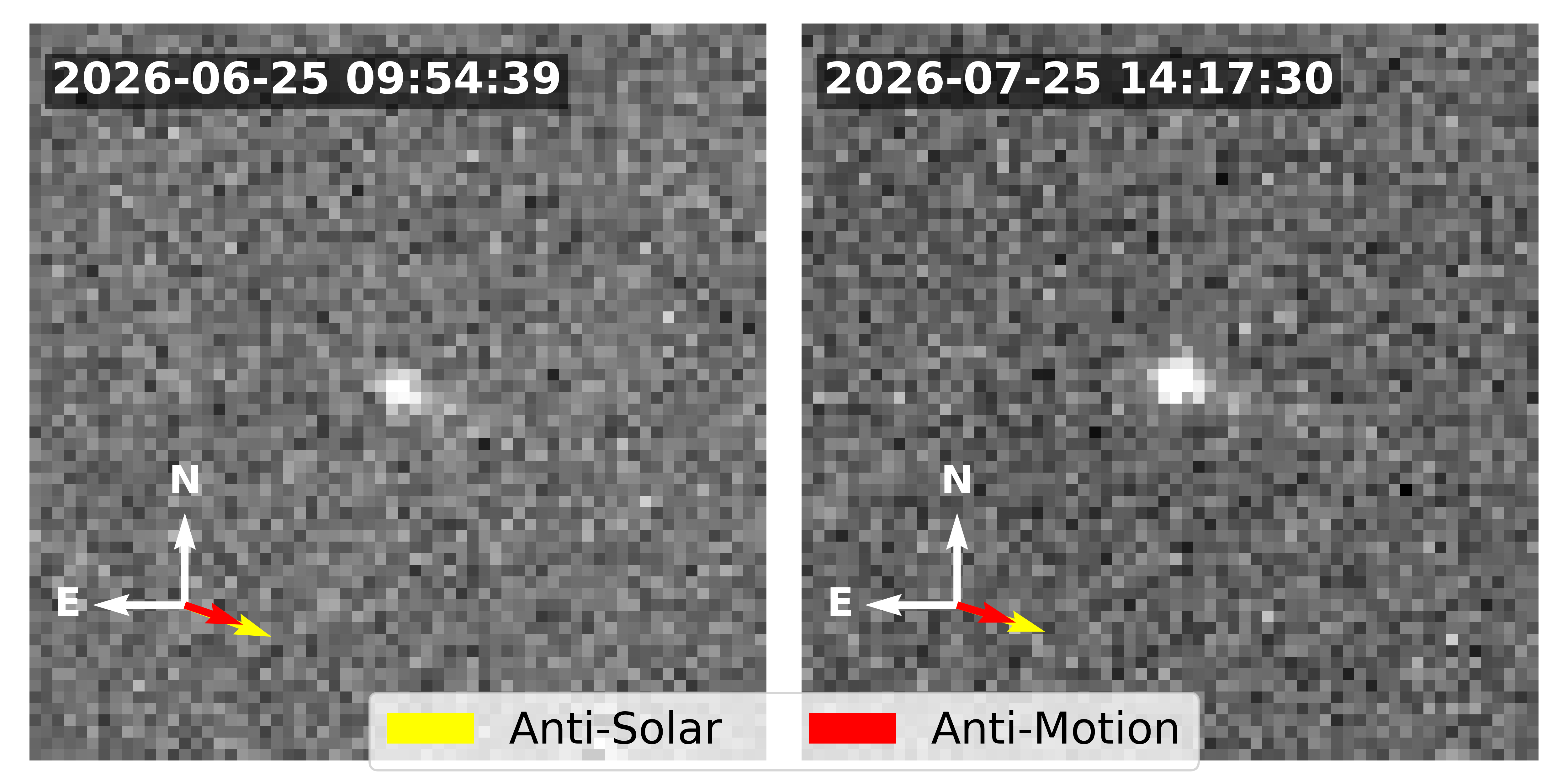}
    
    \caption{(Top) ZTF alert stamps for asteroid 6993 from UT 2026~June~25 through UT 2026~August~5. The exposure on the right on UT 2026~August~5 was the one identified by our CNN. The ZTF pixel scale is 1.0 arcseconds/pixel. (Bottom) ATLAS Mauna Loa Observatory exposures for asteroid 6993 on UT 2026~June~25 and UT 2026~July~25. Images were taken with the 0.5 m Wright-Schmidt Telescope, with a pixel scale of 1.860 arcseconds/pixel. (Both) Anti-solar (yellow) and anti-velocity (red) position angles overlap and align with the direction of the visible tail in each exposure, oriented southwest. The directions of North and East are indicated in each panel by white arrows and the position angles extend at arbitrary lengths, to show that they overlap.
}
    \label{fig:2026_stamps}
\end{figure}

Our CNN identified a 
ZTF observation of asteroid 6993 on UT 2026~August~5
as active at 99.9\% confidence; this score was among the top~10 most probable active asteroid candidates out of 63,974~images\footnote{These observations were accessed through the SNAPS web portal at {\url{http://snaps.nau.edu/}
}} analyzed from that night. Visible activity can be seen in earlier ZTF data, on nights that the CNN had not been implemented, starting on UT 2026~June~25 (Figure\,\ref{fig:2026_stamps}, top). Each exposure reveals a clearly visible tail oriented along the overlapping anti-solar and anti-motion position angles at (252 $\pm$ 2$^\circ$) East of North. The narrow, straight tail extends at least 32 arcseconds beyond the nucleus in the ZTF exposures. Observations retrieved from the Asteroid Terrestrial-impact Last Alert System (ATLAS; \citealt{Tonry_2018})
 forced photometry server\footnote{\url{https://fallingstar-data.com/forcedphot/}} reveal a visible tail oriented in the same direction as seen in ZTF data, beginning on UT 2026~June~25 (Figure\,\ref{fig:2026_stamps}, bottom). On UT 2026~June~25, 6993 was at true anomaly $\nu=$ 319.1$^\circ$, heliocentric distance $r=$ 2.91 au, and geocentric distance $\Delta=$ 3.36 au. The last ZTF observation of this object prior to UT 2026~June~25 was UT 2025~July~20, in which there is no visible activity, suggesting that this activity event began within the last year. However, we note that techniques such as deep stacking have not yet been performed to identify potential faint activity signatures prior to this activity epoch.

\begin{figure}
    \centering
    \includegraphics[width=0.75\linewidth]{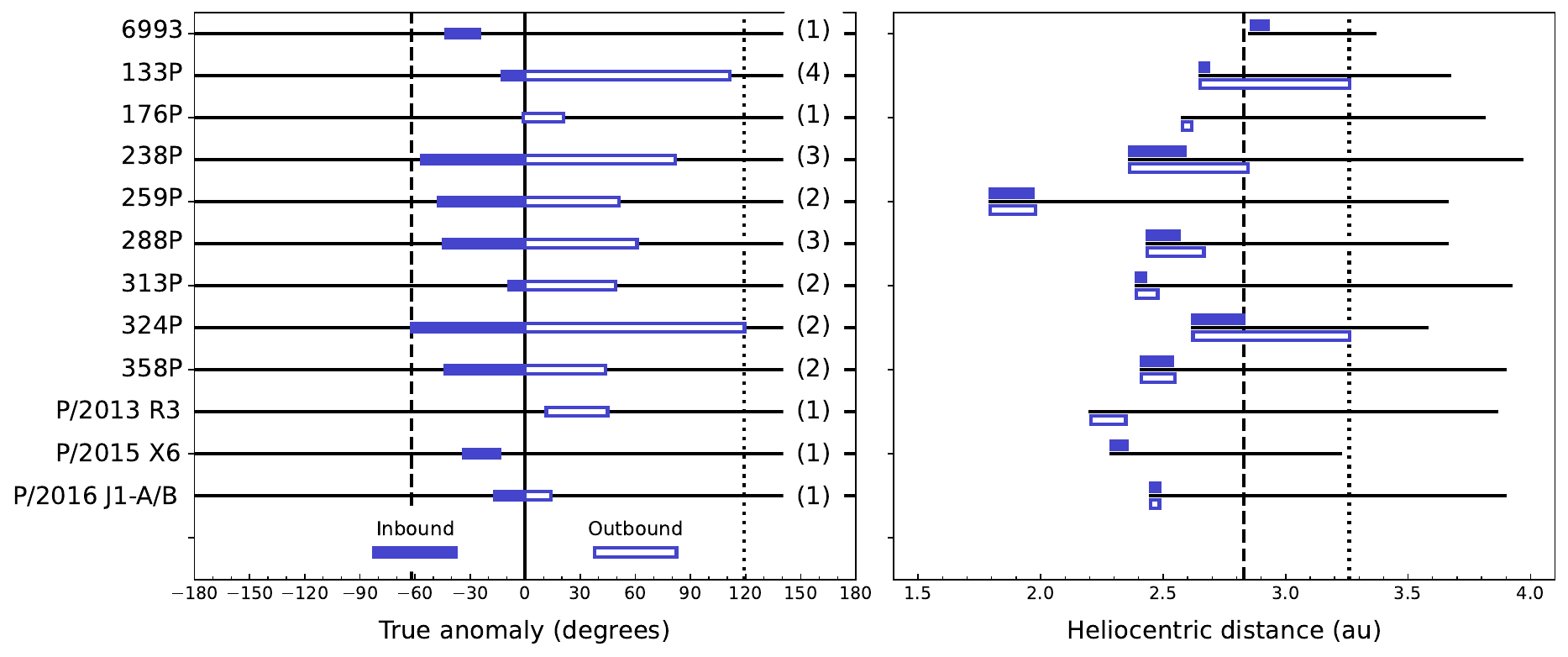}
    \caption{Observed active ranges for known/suspected MBCs appearing in \textcite{Hsieh_2018} plotted with asteroid 6993's current activity period over true anomaly (left) and heliocentric distance (right). Solid blue boxes are pre-perihelion and the outlined blue boxes are post-perihelion. Numbers in parentheses to the far right of the left panel indicate the number of confirmed active apparitions that have been observed for each object. \textbf{(Left)} The black vertical line in the left plot shows perihelion. Relative to perihelion passage, the earliest and latest observed activity positions on an MBC (324P) are plotted as the vertical dashed and dotted lines, respectively. \textbf{(Right)} Horizontal black lines are the heliocentric distance range in a single-orbit for each object, with the most distant positions in which activity has been observed on an MBC (324P) pre-perihelion (dashed-line) and post-perihelion (dotted-line).  The discovery presented here means that 6993 has a larger pre-perihelion activity detection than these known/suspected MBCs. Note that this epoch of activity for 6993 is ongoing, and thus incomplete on the plot.}
    \label{fig:true_anom_activity}
\end{figure}

\begin{figure}
    \centering
    \includegraphics[width=0.75\linewidth]{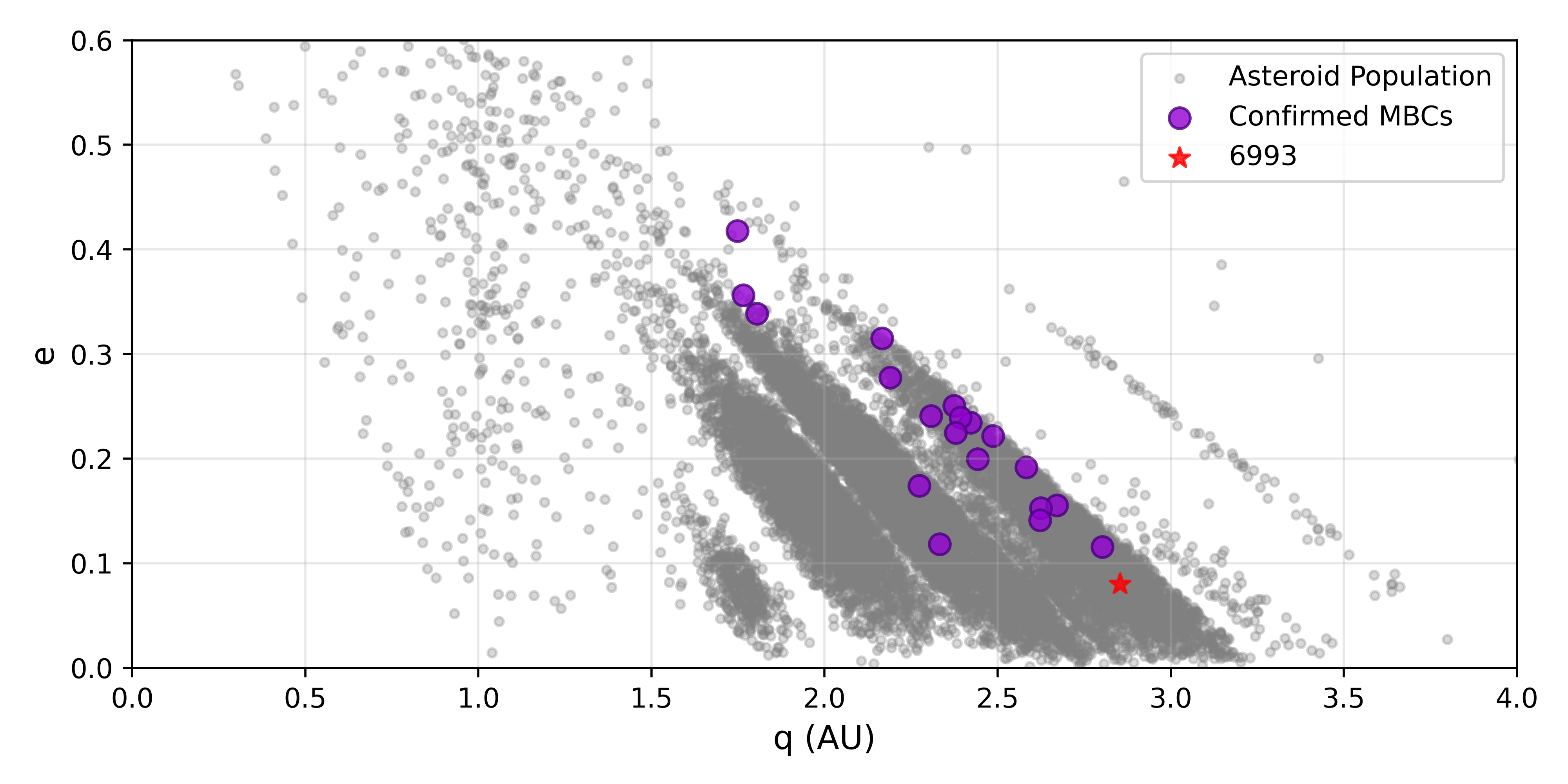}
    \caption{Eccentricity over perihelion distance for 10,000 background asteroids from the Minor Planet Center\footnote{See https://minorplanetcenter.net/data} (grey) along with confirmed MBCs (purple). The red star is asteroid 6993.}
    \label{fig:e_vs_q}
\end{figure}

Traditionally, sublimation driven activity is associated with activity periods observed during multiple perihelion passages. With an orbital period of 5.48 years, there have been only two
near-perihelion epochs in which ZTF observed this asteroid. ZTF
observed asteroid 6993 from 2021~January to 2021~February until true anomaly $\nu=$ 328$^\circ$, but no tail is visible in the data from that season.
Similarly, in archival ATLAS data from early 2021, the object shows no signs of visible activity. Evolving activity strength across perihelion passages has been reported for MBCs 238P and 288P \citep{Hsieh_2018}. It is possible that activity strength for 6993 has increased over time, like 288P, thus making it detectable by 0.5--1.2~m apertures. It is also possible that if this activity episode observed from UT~2026~June~25 through 2026~September~22 is sublimation driven, the object became visibly active after the ZTF and ATLAS 2021 near-perihelion observations ceased.

\section{Discussion}
We compare the current activity event observed for asteroid 6993 with
a collection of sublimation driven activity epochs for a group of
suspected/known MBCs in Figure~\ref{fig:true_anom_activity}. The observed activity detection for 6993 at $\nu=$ 319.1$^\circ$ (-40.9$^\circ$ before perihelion) occurs well within the range observed activity for the known/suspected MBCs appearing in \textcite{Hsieh_2018}. However, until now, sublimation driven activity has yet to be observed on an inbound object at heliocentric distances greater than 2.64~au. Figure~\ref{fig:e_vs_q} shows the distribution of confirmed MBCs in eccentricity and perihelion distance, along with asteroid 6993. Asteroid 6993 has a similar semi-major axis but more circular orbit than most MBCs. If this activity is sublimation driven, asteroid 6993 would have the furthest perihelion distance (q=2.85 au) out of all of the MBCs. Assuming a bond albedo of 0.09, emissivity of 0.9, an effective spherical shape with diameter D = 11.85 km, and even temperature distribution across the surface, we estimate an equilibrium surface temperature of 231~K at aphelion, 251~K at perihelion and 248~K at the time of first observed activity on 25~June~2026. Compared to aphelion, the solar flux is 39$\%$ more intense at perihelion. This narrow range of temperatures, and possibly coldest perihelion temperature of all known MBCs, may both tightly constrain the distribution of potential volatiles on 6993 and in the main belt, and further detailed assessment of its activity is warranted.
This asteroid will reach perihelion in 2027~January,
and observations near and following perihelion passage may reveal greater mass flux as a function of heliocentric distance, which would support sublimation driven activity.


\textcite{Carvano_2010} classify asteroid 6993 as a D-type in their SDSS-color based taxonomic classification.  The D-type taxonomy is typically associated with dark organic-icy materials while their spectra resemble comet nuclei and are associated with outer Solar System bodies \citep{Hartmann_1987, Fitz_1994, Lamy_2004, Nuth_2020}. However, the comparatively high NEOWISE geometric albedo of 0.166$\pm$0.030; \citealt{Masiero_2012} likely contradicts the D-type complex, and is dissimilar to measured albedos for MBCs 133P/Elst-Pizarro and 176P/LINEAR at (0.05$\pm$0.02) and (0.06$\pm$0.02), respectively \citep{Hsieh_2009}. In fact, other MBCs have been classified as F/B/C-types \citep{Bagnulo_2010, Licandro_2011, Licandro_2013}, rather than D-types. Over the 463 observations from 2019~May~4 to 2025~July~20 (prior to this activity epoch), SNAPS derives a mean observed (including the solar color) $g-r$ and $r-i$ (ZTF $g$, ZTF $r$, and ZTF $i$) color of 0.57$\pm$0.04 and 0.28$\pm$0.03 respectively, which is slightly
red among the suite of SNAPS-derived asteroid colors \citep{SNAPS}.
These colors agree with the SDSS observed Sloan $g-r$ and Sloan $r-i$ colors of 0.61$\pm$0.04 and 0.30$\pm$0.04, respectively\footnote{Derived from the 4th Release of the Sloan Digital Sky Survey Moving Object Catalog found at \url{https://faculty.washington.edu/ivezic/sdssmoc/sdssmoc.html}}. We derive ZTF $g-r$ and $r-i$ observed colors of 0.64$\pm$0.02 and 0.19$\pm$0.02 over the 65 observations during this current activity epoch, which are marginally different. Both the SDSS and the two epochs of ZTF observed colors are all broadly consistent with, but at the edge of, the color range for D-type asteroids in \cite{Carvano_2010}. However, they are also consistent with silicate type ranges (S, Q). Given the NEOWISE albedo and colors, a silicate-type composition may be more likely. Additionally, using the HG system \citep{Bowell_1989}, we derive the slope parameter G in the ZTF $r$ filter as 0.22$\pm$0.03. This places asteroid~6993 intermediate between the C and S complex populations in the space defined by G$_r$ and $g-r$ color in published SNAPS data \citep{SNAPS}, which does not help resolve the uncertainty about the taxonomic classification for asteroid~6993.
In summary,
asteroid 6993 shares some characteristics with MBCs dynamically, but would be an outlier compositionally, 
depending on the degree of apparent contradiction between its color and asteroid taxonomy and its relatively high albedo. 

\begin{figure}[ht!]
\centering
    \includegraphics[width=0.40\textwidth]{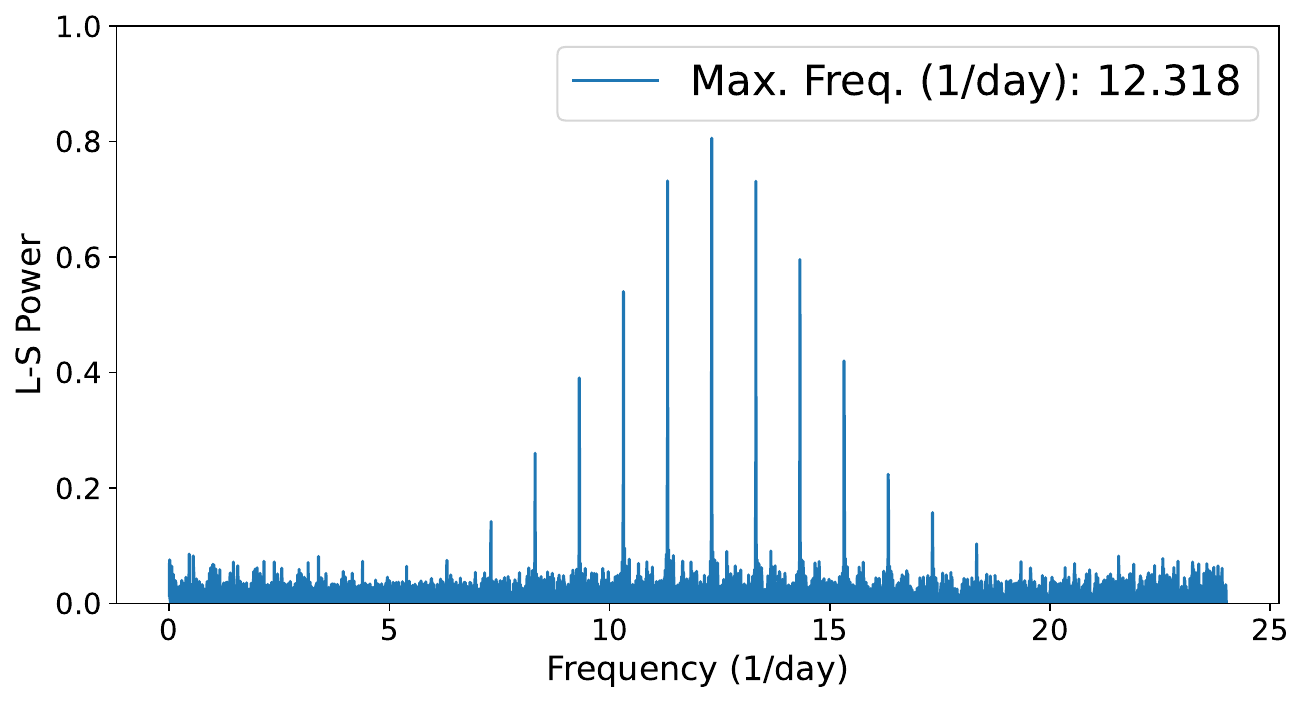}
    \includegraphics[width=0.5\textwidth]{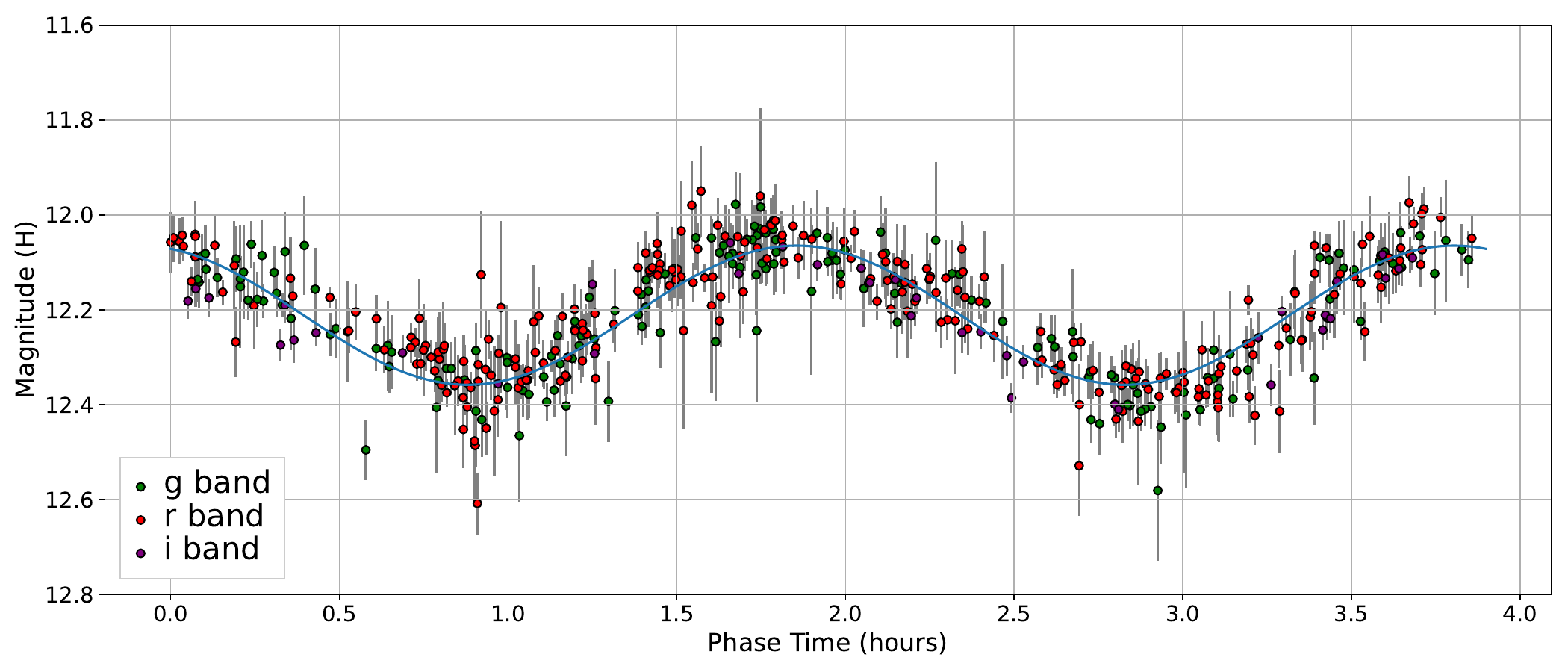}
    
    \caption{(\textbf{Left}) Lomb-Scargle power as a function of frequency of the lightcurve solution, shown in inverse days, for observations from UT 2019~May~4 to 2025~July~20. The power peaks at a light curve period of 0.0812~days, which is equivalent to a rotation period (which is twice the lightcurve period) of 3.897~hours. (\textbf{Right}) Photometric data for~6993 phased to the rotation period of 3.897~hours and amplitude of 0.29~mag. Observations in the ZTF $g$ (green), ZTF $r$ (red), and ZTF $i$ (purple) are corrected to the ZTF $r$ filter in order to derive $g-r$ and $r-i$ colors of 
    0.57$\pm$0.04 and 
    0.28$\pm$0.03, respectively. 
}
    \label{fig:light_curve}
\end{figure}

We also consider the possibility that
the activity seen emitting from asteroid 6993 is due to rotational instability.
Among the derived properties SNAPS produces are rotation period, amplitude, and absolute magnitude ($H$). No previous lightcurve properties were published for this object;
the best-fit Lomb-Scargle period solution found by SNAPS
(as described in \citealt{SNAPS}) for the 463~observations prior to this activity epoch is 3.897~hours with an amplitude of 0.29~mag (Figure~\ref{fig:light_curve}).
This means asteroid 6993 is not rotating at or faster than the spin barrier (generally a rotation limit around 2.2~hours for bodies larger than 200~m; \citealt{Pravec_Harris_2000, Warner_2009}). While it is unlikely that rotational instability is driving this activity, it is possible that the rotation rate could be contributing to activity. This kind of contribution is hypothesized for 133P \citep{Jewitt_2014}, whose rotation period (3.47 hours) is comparable with the value derived for 6993.
We note that 6993 is also twice the size of the larger MBCs (D$<$5.1~km; \citealt{Hsieh_2023}), and objects of this size are less affected by the Yarkovsky–O’Keefe–Radzievskii–Paddack (YORP; \citealt{Rubincam_2000, Vokrouhlick_2015}) effect, which becomes significant for main-belt asteroids at D$\lesssim$5~km.

\begin{figure}[ht!]
\centering
    \includegraphics[width=0.45\textwidth]{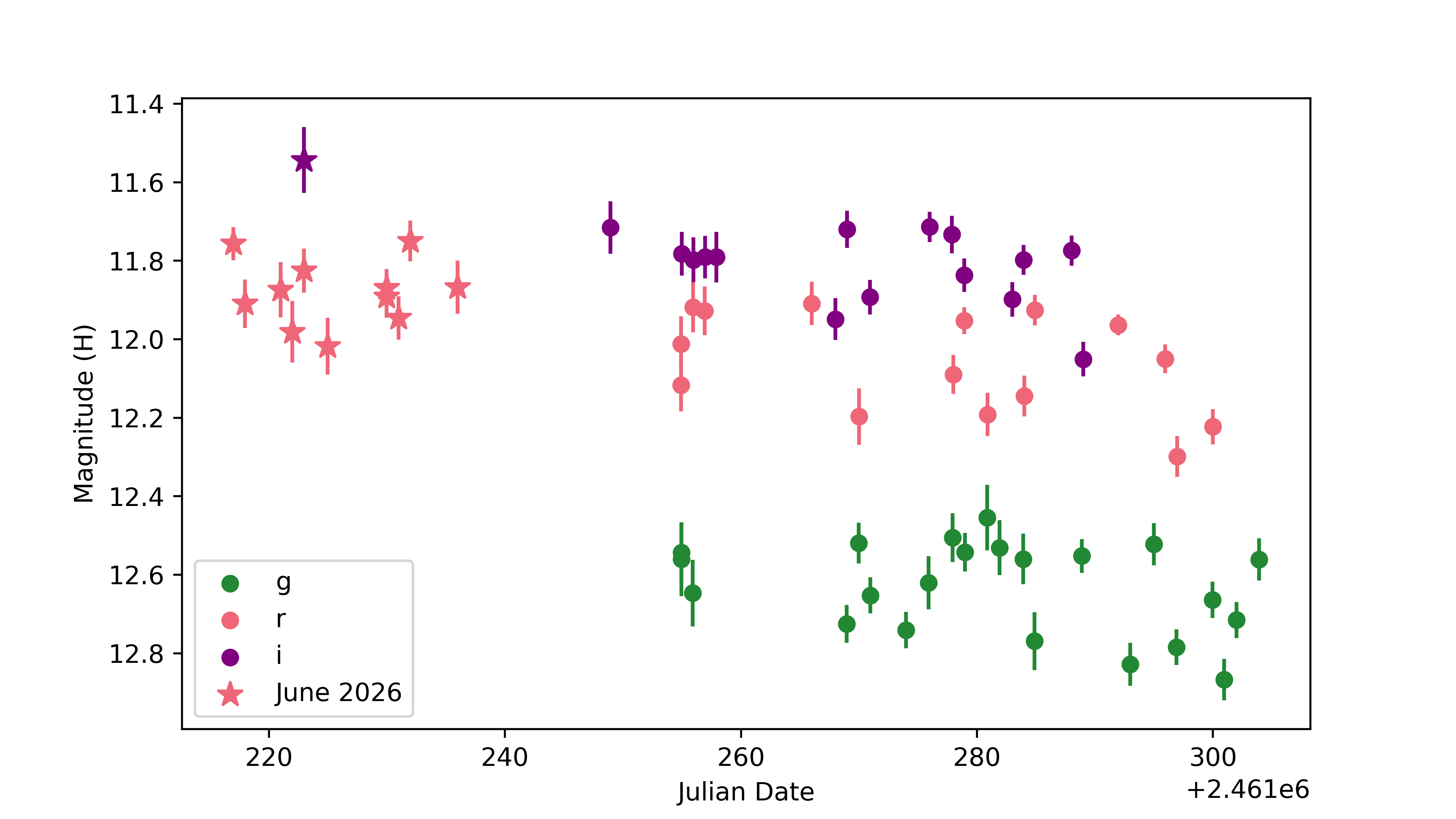}
    \hspace{0.3cm}
    \includegraphics[width=0.45\textwidth]{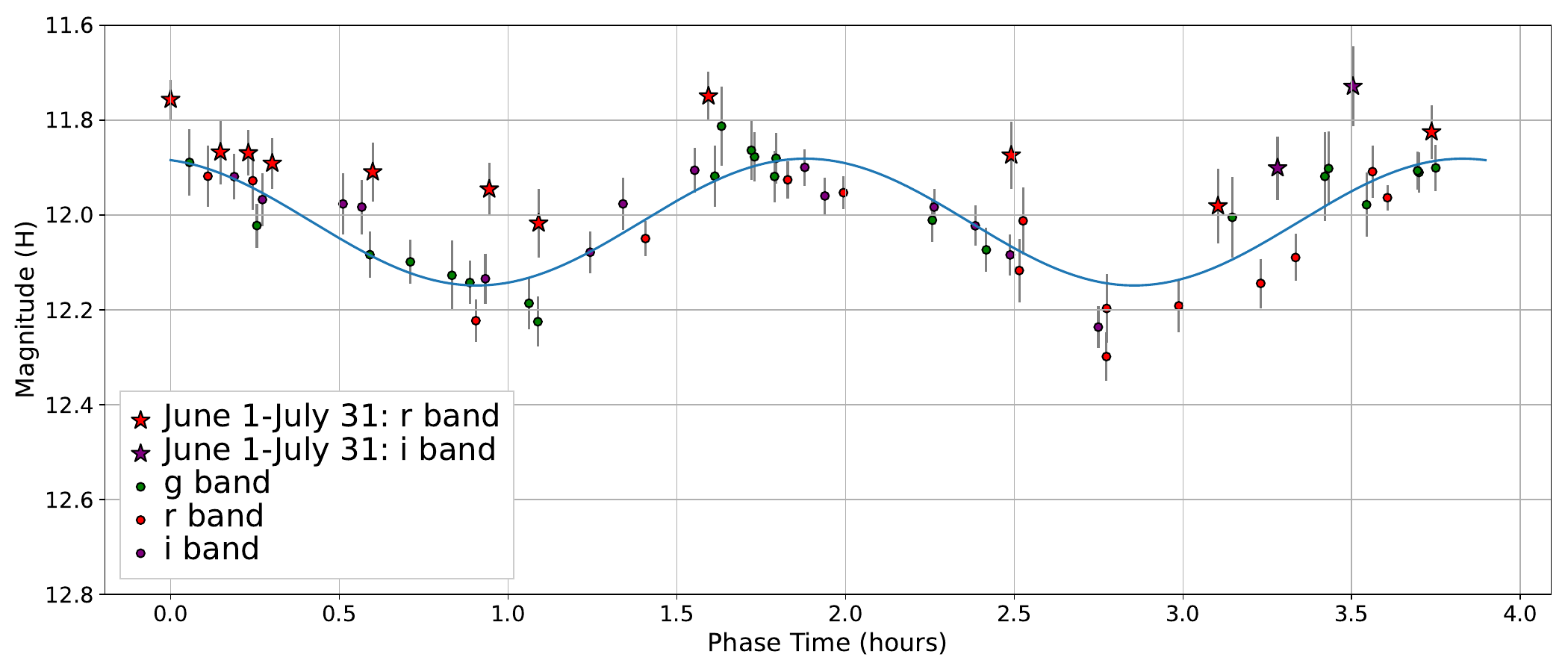}
    
    \caption{(Left) Absolute magnitude ($H$) in ZTF $g$, ZTF $r$, and ZTF $i$ filters for 6993 from UT 2026~June~25 through  2026~September~22. The object has a visible tail during all 65 observations in this range. The stars represent observations during June~2026. From  2026~June~25 to 2026~September~22, 6993 decreases in brightness by $\sim$0.20 magnitudes. (Right) The same observations shown in the left figure, fit to a derived rotation period of 3.896 hours and amplitude of 0.27. We derive $g-r$ and $r-i$ colors of 
    0.64$\pm$0.02 and 
    0.19$\pm$0.02, respectively during this activity epoch. The stars represent the observations from UT 2026~June~01 to 2026~July~31.
}
    \label{fig:6993_vs_time}
\end{figure}

It is also possible that this activity is a result of an impact event. ZTF observations from UT 2026~June~25 to 2026~September~22 are shown in Figure~\ref{fig:6993_vs_time}, all of which are during this current activity epoch. The June-July observations are $\sim$0.2 magnitudes brighter than observations in August and September. It is possible that an impact event occurred sometime around or before UT 2026~June~25, resulting in excavated dust close to the nucleus and thus a brighter apparent magnitude. In this scenario, by August and September the dust would have dissipated from the nucleus and therefore the apparent magnitude would be somewhat fainter. 
This scenario would be supported if near-future observations show a continued decrease in brightness, and eventual return to pre-outburst levels.
In addition, if this is a silicate-type composition, the likelihood of an impact scenario rather than sublimation is higher. Observations through the next months as it passes perihelion and into the next orbit (i.e., in a search for repeated activity near perihelion) are warranted to better characterize this activity.

\section{Conclusions and Future Work}
We report the discovery of activity on outer-main belt asteroid 6993 and confirm its designation as an active asteroid with $T_J=$ 3.2. The object was flagged by our SNAPS machine learning model at 99.9$\%$ confidence just hours after the discovery image was ingested on UT 2026~August~5, and confirmed on multiple ZTF observations from UT 2026~June~25 through 2026~September~22. Visible activity is also observed in ATLAS data as early as UT 2026~June~25. This object appears to have become active at true anomaly $\nu=$ 319.1$^\circ$, a similar 
position to the first activity detection range on other MBCs. Archival ZTF and ATLAS data do not resolve any significant emission around its previous perihelion passage, making it challenging to comment on the recurrent nature of this activity. So far, this object is consistent with the MBC classification dynamically, but outlying compositionally and in size. The taxonomic classification and measured albedo are not in agreement, which highlights the need for further characterization of
this object. We also find plausible evidence for an impact  based on an observed decrease in brightness since July. It is also possible that the activity is the result of multiple mechanisms (i.e., rotation and sublimation, impact and sublimation).
Further telescope follow up should be carried out to monitor the brightness and morphology of the tail as a function of heliocentric distance to distinguish between activity mechanisms. For example, development of another tail may indicate long duration dust emission \citep{Hsieh_2012}; long sustained activity with outbursts continuing out to aphelion may indicate rotational instability; and gradual tail dissipation leading up to perihelion could support the impact argument. The low eccentricity of 6993's orbit compared to other MBCs motivates understanding its potential volatile driven activity, which would help inform the volatile inventory in the asteroid belt. Alternatively, z-band measurements may reveal a silicate-rich composition, which would further support the impact argument.

ZTF and LSST are both producing rich data streams that will enable the detection of asteroid activity; this discovery, just hours after the observation on UT 2026~August~5, demonstrates our readiness to make those detections essentially in real time.
At full scale,
ZTF and LSST yield $\sim$10$^4$ and $\sim$10$^5$ asteroid observations per night, respectively. Considering the current occurrence rate estimates for active asteroids ($\sim$1 in 10,000), LSST may observe anywhere from 1--100~active asteroids on a given night. This presents an opportunity to increase the sample of known active asteroids to enable population studies of the activity driving mechanisms and constrain the abundance of water in the asteroid belt.

\section{Acknowledgments}
This work was funded in part by NSF 
AAG award
\#2206796,
NASA SSO award \#80NSSC25K7728, and the Northern Arizona University division of the Arizona Space Grant Consortium through NASA grant \#80NSSC25M7084.
Finally, this research has made use of data and/or services provided by the International Astronomical Union's Minor Planet Center. We thank the referee for useful comments that improved this paper.

This work is based in part on observations obtained with the Samuel Oschin Telescope 48-inch and the 60-inch Telescope at the Palomar Observatory as part of the Zwicky Transient Facility project. ZTF is supported by the National Science Foundation under Grants No.\ AST-1440341 and AST-2034437 and a collaboration including current partners Caltech, IPAC, the Oskar Klein Center at Stockholm University, the University of Maryland, University of California, Berkeley, the University of Wisconsin at Milwaukee, University of Warwick, Ruhr University, Cornell University, Northwestern University and Drexel University. Operations are conducted by COO, IPAC, and UW.

This work includes measurements from the Sloan Digital Sky Survey.
Funding for the SDSS and SDSS-II has been provided by the Alfred P. Sloan Foundation, the Participating Institutions, the National Science Foundation, the U.S. Department of Energy, the National Aeronautics and Space Administration, the Japanese Monbukagakusho, the Max Planck Society, and the Higher Education Funding Council for England. The SDSS Web Site is http://www.sdss.org/.

The SDSS is managed by the Astrophysical Research Consortium for the Participating Institutions. The Participating Institutions are the American Museum of Natural History, Astrophysical Institute Potsdam, University of Basel, University of Cambridge, Case Western Reserve University, University of Chicago, Drexel University, Fermilab, the Institute for Advanced Study, the Japan Participation Group, Johns Hopkins University, the Joint Institute for Nuclear Astrophysics, the Kavli Institute for Particle Astrophysics and Cosmology, the Korean Scientist Group, the Chinese Academy of Sciences (LAMOST), Los Alamos National Laboratory, the Max-Planck-Institute for Astronomy (MPIA), the Max-Planck-Institute for Astrophysics (MPA), New Mexico State University, Ohio State University, University of Pittsburgh, University of Portsmouth, Princeton University, the United States Naval Observatory, and the University of Washington.

This work has made use of data from the Asteroid Terrestrial-impact Last Alert System (ATLAS) project. The Asteroid Terrestrial-impact Last
Alert System (ATLAS) project is primarily funded to
search for near earth asteroids through NASA grants
NN12AR55G, 80NSSC18K0284, and 80NSSC18K1575;
byproducts of the NEO search include images and catalogs from the survey area. This work was partially
funded by Kepler/K2 grant J1944/80NSSC19K0112 and
HST GO-15889, and STFC grants ST/T000198/1 and
ST/S006109/1. The ATLAS science products have been
made possible through the contributions of the University of Hawaii Institute for Astronomy, the Queen’s University Belfast, the Space Telescope Science Institute,
the South African Astronomical Observatory (SAAO),
and The Millennium Institute of Astrophysics (MAS),
Chile.

\bibliography{sample701}{}
\bibliographystyle{aasjournalv7}



\end{document}